\documentclass[12pt]{article}
\usepackage{amsmath,amssymb,times,graphicx}
\begin{document}
\pagestyle{plain}
\hsize = 6. in 				
\vsize = 8.5 in		   
\hoffset = -0.3 in
\voffset = -0.5 in
\baselineskip = 0.26 in	

\def\vF{{\bf F}}
\def\vJ{{\bf J}}
\def\vX{{\bf X}}
\def\tf{{\widetilde{f}}}
\def\vv{{\bf{v}}}
\def\tv{{\widetilde{v}}}
\def\tu{{\widetilde{u}}}
\def\tp{{\widetilde{\rho}}}
\def\vx{\mbox{\boldmath$x$}}
\def\tK{{\widetilde{K}}}

\title{Hill's Small Systems Nanothermodynamics: A
Simple Macromolecular Partition Problem with 
a Statistical Perspective}

\author{Hong Qian\\
Department of Applied Mathematics, University of Washington\\
Seattle, WA 98195-2420, U.S.A.}

\maketitle

\begin{abstract}
Using a simple example of biological macromolecules
which are partitioned between bulk solution and 
membrane, we investigate T.L. Hill's phenomenological
nanothermodynamics for small systems.  By introducing
a {\em systems size dependent} equilibrium constant 
for the bulk-membrane partition, we obtain Hill's results
on differential and integral chemical potentials
$\mu$ and $\hat{\mu}$ from computations based on standard 
Gibbsian equilibrium statistical mechanics.  
It is shown that their difference can be understood from
an equilibrium re-partitioning between bulk and 
membrane fractions upon a change in system's size;
it is closely related to the system's fluctuations and
inhomogeneity. These results provide a better understanding
of the nanothermodynamics and clarify its logical relation
with the theory of statistical mechanics.  
\end{abstract}

\section{Introduction}

	With great advances in material preparations at nanoscale
in recent years, there is a growing interest in the 
thermodynamics of small systems \cite{bustamente,abe_review}.  Nanothermodynamics, a 
phenomenological theory developed by T.L. Hill in the
early 1960s for small equilibrium systems, has become 
one of the major formalisms for quantitative treatment
of equilibrium nanoscale materials \cite{hill_nano,hill_book}.  
There are many experimental and computational work
that have verified the theory, see for examples
\cite{chamberlin,kjelstrup}.

The greatest strength of thermodynamics, being
absolute and abstract, is often its weakness: It
usually provides no molecular mechanism and insights
into chemical and biochemical processes.  The
same dichotomy applies to Hill's nanothermodynamics:
The key concept in the theory is a difference
between {\em differential} and an {\em integral} 
froms of many non-extensive thermodynamic quantities 
due to smallness of a system.  For example, since
Gibbs free energy $G(N,p,T)$ is not linearly 
proportional to the systems size $N$, 
$\mu=\partial G/\partial N$ and 
$\hat{\mu}=G/N$ are different.  The difference
is intuitively understood: It is due to contributions
such as surface effect.

In biophysical chemistry, biological macromolecules
are often treated as having discrete conformational
states, and interaction with ligands as stoichiometric
binding.  While this approach is only an approximation, 
it often provides much more powerful intuitive
understanding of the physical processes that
underly molecular thermodynamics.  A case in point 
is J.A. Schellman's theory of three-component
system with mixed solvents \cite{timasheff}.
While a thorough treatment of this problem 
involves perferential interaction coefficients,
discrete stoichiometric binding model has provided
greater intuition \cite{schellman}.\footnote{A difference
between differential and integral forms of
macromolecular interaction also appears in
the theory of binding: Thermodynamic binding 
can go to zero while molecular interaction is the  
strongest \cite{jas_bpc_93}.
}   

	The fundamental premises of a {\em small system} 
is that its thermodynamic properties are not 
strictly proportional to the system's 
size \cite{hill_book}.  A system's size, however,
can be represented by several different quantities:
The number of molecules $N$ in a system; the volume 
$V$ of a system, or the total energy $E$ of a 
system.  In chemistry and biochemistry,  
concentration $x=N/V$ is a widely used intensive 
quantity. We shall refer to system's size in term
of its volume in the rest of the paper.  In 
terms of $V$, the Helmholtz free energy $F$ of
a three-dimensional small system could have the form
\begin{equation}
           F  = Vf(x,T) + a(x,T)V^{2/3} + b(x,T)\ln V 
        + \cdots,
\end{equation}
where the term in $V^{2/3}$ is associated with a
surface energy.

	In the present work, we first
consider the following simple problem: 
$N$ non-interactive macromolecule M are
in a system which consists of bulk solution
and membrane. (One could think of a 
vesicle with lipid membrane.)  
Each M molecule has three discrete
conformational states: 0, 1 in the bulk
and 2 on the membrane.  We assume the
equilibrium constant between states 0
and 1 is $K_1$ and between states 0
and 2 is $K_2$.  $K_1=p_1/p_0$ and 
$K_2=p_2/p_0$ where
$p_i$ is the probability of an
$M$ being in state $i$ in equilibrium.
$K_1$ is independent of amount of membrane
in the system; but $K_2$ is critically
dependent on the ratio of membrane surface
$S$ to bulk volume $V$, thus the system's size.
Assuming simple geometry with surface to 
volume ratio $S/V=V^{-1/3}$ and homogeneity
in both bulk and membrane phases,
it is reasonable to assume that $K_2(V)\propto
V^{-1/3}$.  Thus the equilibrium partition
function for a single M molecule, with the state 0 
as the refererence state:
\begin{equation}
      Z_1 = 1+ K_1 + K_2(V) \rightarrow 1 + K_1
\label{eq_002}  
\end{equation}
in the limit of $V\rightarrow\infty$.
The simple example can be easily generalized to
more general macromolecular partition problem, and
a key formula relating $\mu$ and $\hat{\mu}$ first
appeared in \cite{hill_book}, will be derived.  

A few remarks on Eq. (\ref{eq_002}) are in order:
($i$) $Z_1$ looks like a binding polynomial, but 
it is actually independent of the concentration of the
M.  This is because we have assumed that the$N$ macromolecules
are non-interactive.  Hence, $Z_1$ is the ``partition function''
of a single M molecule.  The partition function for the
whole system is $Z_N = Z_1^N$. 

($ii$) $K_2(V)=p_2/p_0$ as a function of $V$ is itself a complex
problem in general.  For our simple model, we shall completely
neglect surface free energy of the membrane. Furthermore, if one treats  macromolecule-membrane association as simple Langmuir absorption, with $N(p_0+p_1)/V$ and $Np_2/S$ being the concentrations of the M in bulk and on membrane.  Then one obtains 
$p_2/(p_0+p_1)\propto S/V$.  Since $(p_0+p_1)=(1+K_1)p_0$,
one has $p_2/p_0 \propto S/V \sim V^{-1/3}$.   For the 
present analysis, the key is that $K_2(V)$ is a function of
$V$.  The result in Eq. (\ref{eq_13}) given below does
not depend on a specific functional form of $K_2(V)$.

The novel idea of the present work is to introduce 
the {\em systems size dependent} equilibrium constant(s), 
(i.e., more generally potential of mean force or 
conditional free energy).  It is shown that if one 
starts with a mechanistic based statistical approach, 
following Gibbs, to small systems, then the small 
systems nanothermodynamics naturally emerges.   

	One of the salient features of Hill's 
nanothermodynamics is its {\em ensemble dependence}: 
When a systems boundary becomes significant, how 
a statistical thermodynamic system is maintained at the
boundary matters.  In the past, dependence upon details 
of an ensemble at its boundary leads to the phenomenon 
of {\em entropy-enthalpy compensation} 
\cite{qian_jjh_96,qian_jcp_98}, which can be understood
from relaxing certain internal constraints.  
In Sec. \ref{sec:eec} we show that this perspective 
is also fruitful in understanding the difference 
between differential and integral chemical potentials
$\mu$ and $\hat{\mu}$.

\section{Bulk-Membrane Partition: A Simple Example}

	We now consider a simple system in which there are
$N$ non-interactive macromolecle M.  Each M has three distinct 
states: states 0 and 1 are in bulk solution and state 2 is
associated with membrane.  It is clear that the ratio 
between total membrane to total bulk volume is a function
of the systems size $V$: Thus by simple geometric consideration
$K_2(V)$ is volume dependent.  The partition function for the 
small system with constant pressure $p$ and temperature $T$ is:
\begin{equation}
            Z(N,p,T) = \left(1+K_1+K_2(V)\right)^N,
\label{eq_02}
\end{equation}
in which $V=V(N,p,T)$ is a function of $N$.\footnote{Mathematically
more rigorous, one needs to solve the problem with self-consistency
among Eqs. (\ref{eq_02}), (\ref{eq_03}) and:
$V(N,p,T)=\left(\partial G/\partial p\right)_{N,T}$. 
}
Therefore,
\begin{eqnarray}
	G(N,p,T) &=& -k_BT\ln Z(N,p,T),
\label{eq_03} 
\\[8pt]
    \hat{\mu}(N,p,T) &=& \frac{G(N,p,T)}{N}
               =-k_BT\ln\left(1+K_1+K_2\right),
\label{eq_04}
\\[6pt]
    \mu(N,p,T) &=& \frac{\partial G(N,p,T)}{\partial N}
         = \hat{\mu} -\frac{k_BT K_2
           \left(\frac{\partial\ln K_2}{\partial\ln N}\right)_{p,T}
           }{1+K_1+K_2},
\label{eq_13}
\end{eqnarray} 
where $\hat{\mu}$ and $\mu$ are integral and 
differential chemcial potentials introduced by
Hill \cite{hill_book}.   In the last term 
of Eq. (\ref{eq_13}), 
$\left(\partial\ln K_2/\partial\ln N\right)$
is an intensive quantity, $K_2\propto V^{-1/3}$ and
$1+K_1+K_2 \rightarrow 1+K_1$.  Hence it is 
on the order of $K^{-1/3}$.  Thus, 
$\mu$ and $\hat{\mu}$ are the same
for a macroscopic system when  $G(N,p,T)$ 
asymptotically becomes an ``extensive'' quantity: 
$G\propto N$.

Eq. (\ref{eq_13}) can be easily generalized
into
\begin{equation}
   \mu(N,p,T)-\hat{\mu}(N,p,T) = \Big\langle
          \left(\frac{\partial \mu_i^o(N,p,T)}{\partial\ln N}  
             \right)_{p,T}\Big\rangle =
           \left(\frac{\partial\hat{\mu}(N,p,T)}{\partial\ln N}
             \right)_{p,T},
\label{eq_14}
\end{equation}
where 
\begin{equation}
	\mu_i^o(N,p.T)= -k_BT\ln K_i(N,p,T)
\end{equation}
and the average $\langle\cdots\rangle$
is performed with respect to $i=0,1,2$. 
Eq. (\ref{eq_14}) clearly shows why and how the 
small systems thermodynamics depends on the nature of 
ensemble; and why the difference $\mu$ and $\hat{\mu}$
will be ensemble specific.  Eq. (\ref{eq_14})
agrees with what obtained by Hill 
(\cite{hill_book}, Part I, p. 30, Eq. 2-12):
\begin{equation}
    \mu-\hat{\mu} = \left(\frac{\partial\hat{\mu}}{\partial\ln N}
               \right)_{p,T}.
\end{equation}

\section{Thermodynamic Hierarchy and 
Nanothermodynamics}

\subsection{Perturbation and relaxation in
thermodynamic systems}
\label{sec:eec}

The chemical potential $\mu$ is the increment in
free energy of a system when its number of 
molecules increased by 1 \cite{ben-naim_book}. 
Therefore, it can be considered as a {\em
perturbation} to the thermodynamic system.
The idea advanced in \cite{qian_jjh_96}, 
originally proposed by A. Ben-Naim \cite{ben-naim_75},
is to view the response of a thermodynamic
system to a perturbation as two parts:

($i$) Constraining the distribution among all the 
populations un-changed, what is the increase in
free energy of the system due to the perturbation?

($ii$) Inevitably, however, the perturbation will
cause a shift (i.e., re-distribution, re-partitioning) 
in the equilibrium among the populations.  There is an
additional free energy change associated with this
re-organization process.

	This perturbation-relaxation view of thermodynamics 
of complex systems yielded a possible explanation for the
so-called ``entropy-enthalpy compensation'' phenomena
\cite{ben-naim_75,qian_jjh_96,qian_jcp_98}: Note that the 
chemical potentials for all the subcomponents, $\mu_j$, of 
an equilibrium system have to be the same.  Thus, the free 
energy change associated with ($ii$) is essentially zero if
the perturbation is infinitesimal.
In contrast, the corresponding entropy and enthalpy
of the subcomponents, $s_j$ and $h_j$ ($h_j-Ts_j=\mu_j$),
can be very different.  Thus, a large contributions to
entropy and enthalpy from the process in ($ii$) are 
expected.\footnote{Both steps in this gedankenexperiment 
are carried out under an isothermal condition.  They 
can not be realized in a laboratory.  They 
are different from the adiabatic (isoentropic) and
isothermal processes in the derivation of the
fundamental equation of thermodynamics.}  

	Applying the same idea to thermodynamics of our
simple example, a perturbation leads to a shift in equilibrium
between the ``bulk'' and ``membrane'' fractions. 
More specifically, introducing an additional M molecule
causes a change in the system's volume $V$ 
under isobaric (i.e., constant $p$) condition.
This change in $V$ is global: It causes the change of 
$K_2$ for each and every molecule M in the system.

\subsection{Spatial inhomogeneity}

``Size'' is a geometric concept.  The
unique feature of nanothermodynamics of small
systems is {\em spatial inhomogeneity} that 
partitions the particles in a system into 
different populations whose proportions scale 
differently with systems size.  In Hill's theory, this 
partition is implicit: cluster growth naturally
leads ``surface'' and ``interior''; in our
simple model this partition is explicitly 
assumed: states 0 and 1 versus 2.   

	With this geometry in mind, there will be 
two answers to the following question: What
is the change in system's free energy if one
introduces one additional particle into the
system?  Accordingly, answer ($i$) assumes that the 
proprotion is unchanged; and answer ($ii$)
assumes there is a shift in equilibrium 
distributions between different populations.

The answer ($i$) is precisely the macroscopic
result, $\hat{\mu}$.  It is intimately related 
to the free-energy perturbation 
theory \cite{zwanzig,widom}:
\begin{equation}
	\Delta G^o(A\rightarrow B)
    = -k_BT \ln\Big\langle \exp\left(-\frac{E_B-E_A}{k_BT}
               \right)\Big\rangle_A,
\label{eq_10}
\end{equation}
in which the average is carried out in terms of
the equilibrium distribution of the unperturbed
system.  Note that Eq. (\ref{eq_10}) assumes
that the phase space, over which the ensemble average
$\langle\cdots\rangle_A$ is carried out, is the same
before and after the perturbation.

The answer ($ii$) however, contains exactly a contribution 
from small systems effect: The addition of a particle
changes the size of the system, thus the phase space, on 
the order of $1/N$.  This change causes a shift in the equilibrium
between ``surface population'' to ``bulk poplulation'',
which affects all $N$ particles in the system.
This is given in Eq. (\ref{eq_14}).

	Realizing the difference between these two answers
led Hill to introduce $dN_t$ and $d\mathcal{N}$.  The
former corresponds to a change in the total number of 
particles, thus gives $\mu$ (answer $ii$); 
the latter changes the number of particles but keeps
the equilibrium population proportion by ``adding an
independent copy of the small system'', thus gives
$\hat{\mu}$ (answer $i$).

\subsection{Rapidly stirred biochemical
reaction systems}

	For a small biochemical reaction system which is rapidly 
stirred, the theory of Delbr\"{u}ck-Gillespie process
\cite{qian_nonl_11,qian_jsp_10}, with its probability
distribution given by the chemical master equation and
its trajectories follow Gillespie algorithm, predicts 
that its stationary probability distribution, in the limit 
of large size, has the generic form
\begin{equation}
        f(x) \propto \exp\left(-V\phi(x)\right),
\end{equation} 
where $x$ is the concentration of biochemical
species, $V$ is system's volume, and $\phi(x)$ is 
a function of only intensive quantities.  Applying 
Laplace's method for integrals \cite{qian_jmb_06,murray_book}, 
this result leads to the conclusion that for a rapidly 
stirred system without geometric inhomogeneity, the 
partition function 
\begin{eqnarray}
    Q(V,\mu,T) &=& V\int dx\ e^{-V\phi(x)}
\nonumber\\
     &\approx& V e^{-V\phi\left(x^*\right)}
        \int  dx\ e^{-\frac{V}{2}\phi''\left(x^*\right)
                \left(x-x^*\right)^2}
\nonumber\\[6pt]
	&=&  e^{-V\phi\left(x^*\right)}
			\sqrt{\frac{2\pi V}{\phi''\left(x^*\right)}},
\end{eqnarray}
where $x^*(\mu)$ is the macroscopic concentration.
Therefore,
\begin{equation}
   -k_BT\ln Q(V,\mu,T) = k_BTV\phi(x^*) + O(\ln V),
\label{eq_012}
\end{equation}
where $O(\cdots)$ is the mathematical symbol
called Bachmann-Landau notation.  It stands 
for ``on the order of $\cdots$''.  Eq. (\ref{eq_012})
should be compared with Eq. (\ref{eq_02}):
$-k_BT\ln Z(N,p,T)$ $=$ 
$-k_BTVx^*\ln(1+K_1)+O\left(V^{2/3}\right)$ where $x^*=N/V$.
For rapidly stirred nanoscale systems, the correction
to the extensive term is on the order of logarithm of 
an extensive quantity: $\ln V$ or $\ln N$.
When there is a spatial partitioning, e.g., 
compartimentalization, however, terms on the order of 
fractional power of the extensive quantity arise.
For system with bulk-membrane partition,  this term
is on the order of $V^{2/3}$.

	A biological cell is a small thermodynamic system.  Applying 
the present result to a collection of cells, one realizes that 
there are two macroscopic limits: ($a$) Classical biochemistry
studies cell extracts with membrane removed; ($b$) permeable 
cells that preserve the ratio between membrane surface area 
to bulk volume.  They correspond precisely the two ensembles 
of Hill's.

\subsection{Kinetics of re-partitioning}

	All the above discussion has been exclusively 
on equilibrium thermodynamics.  We suspect there is in 
fact a kinetic aspect of the thermodynamics of small 
systems.  Nonequilibrium thermodynamics for
heterogeneous systems and for nanoscale systems
have become matural subjects in recent 
years \cite{NET_book,rubi_jpc,ge_qian_pre_10}.
If one takes a kinetic perspective of the
response of a small system upon perturbation
\cite{qian_jjh_96}, one sees that the 
conformation re-partitioning between the bulk
and surface could indeed be observed
kinetically.  This connection between thermodynamics
and kinetics deserves further investigation
\cite{hill_chamberlin_pnas,rubi_1,rubi_2}.


\section{Acknowledgement}

I thank Prof. Dick Bedeaux for a stimulating week of 
discussions in June 2008 at XXI Sitges Conference (Spain),  
Profs. Ralph Chamberlin and Signe Kjelstrup for 
discussions that stimulated and renewed my interests in nanothermodynamics.
I would like to acknowledge my friend and teacher (via
his writing) Terrell Hill whose work has had major
influence on mine, not only in nanothermodynamics, but
also in nonequilibrium steady state cycle kinetics and
theory of muscle contraction 
\cite{qian_bpc_97,qian_jpcm_05,qian_jpc_06}.

\end{document}